# Direct observation of magneto-electric Aharonov-Bohm effect in moiré-scale quantum paths of minimally twisted bilayer graphene

Yi-Wen Liu, Ya-Ning Ren, Chen-Yue Hao, Lin He[†]
Center for Advanced Quantum Studies, Department of Physics, Beijing Normal University, Beijing,100875, People's Republic of China
[†]Correspondence and requests for materials should be addressed to L.H. (e-mail: helin@bnu.edu.cn).

**Aharonov-Bohm (AB) effect, the well-known archetype of electron-wave interference phenomena, has been explored extensively through transport measurements. However, these techniques lack spatial resolution that would be indispensable for studying the magnetic and electrostatic AB oscillations at the nanometer scale. Here, we demonstrated that scanning tunneling microscopy (STM) can be used as an AB interferometer operating on nanometer length scales and the magneto-electric Aharonov-Bohm effect in minimally twisted bilayer graphene (TBG) was directly measured by using STM. In the minimally TBG, there is a triangular network of chiral one-dimensional states hosted by domain boundaries due to structural reconstruction. Taking advantage of the high spatial resolution of the STM, both the magnetic and electrostatic AB oscillations arising from electron interference along moiré-scale triangular quantum paths in the minimally TBG were measured. Our work enables measure and control of the AB effect and other electron-wave interference at the nanoscale.**

The phase of electron wave packets circling along a closed trajectory can be modulated by electronmagnetic potentials, which is now known as the Aharonov-Bohm (AB) effect[1-11]. As schematically shown in Fig. 1a, one coherent electron beam splits into two moving along two paths and they interfere after recombination. The phase difference between two paths of electron beams is $\Delta\phi = 2\pi(e/h)\Phi$ ($\Phi$ is the magnetic flux encircled by the paths, $h$ is Planck's constant and $e$ is the electron charge), which generates cycles of destructive and constructive interference of the electron wave functions. The phase tuning strongly affects the magnetoresistance in small devices and gives rise to conductance oscillations with the period of one flux quantum $\Phi_0 = h/e$ in transport measurements. Consequently, the AB effect has been studied extensively in transport techniques[2-9]. In the meanwhile, the density of states (DOS) along the paths of electrons are also sensitive to the interference of the wave functions. Scanning tunneling microscopy (STM) is known as a powerful tool to measure the local DOS. Therefore, it was proposed to use STM as an AB interferometer to directly study the AB effect[12,13]. However, such a measurement has so far not been achieved. Here we report on the first measurement of the AB effect by using STM and our experiment demonstrates that the high spatial resolution of the STM enables us to detect both the magnetic and electrostatic AB oscillations in a moiré-scale triangular quantum path in minimally twisted bilayer graphene (TBG).

In the TBG with twist angle $\theta$ smaller than a characteristic angle $\theta_c$ ($\theta_c \sim 1.1°$), there is a strain-accompanied lattice reconstruction that results in large triangular Bernal (*AB* and *BA*) stacking domains and a triangular network of domain walls (DWs)[14-18], as schematically shown in Fig. 1b. When the bandgap is opened in the Bernal-stacking regions, the domain walls can host topological chiral conducting states[19-24], which provide a unique platform to explore the AB effect[25-27]. In a recent transport experiment, giant magnetic AB oscillations were observed in the triangular network of domain walls in the minimally TBG and the main period of the oscillations was attributed to interference in paths encircling a *AB* or *BA* domain[26]. Very recently, analysis based on a phenomenological scattering theory indicated that the main period of the magnetic

AB oscillations should correspond to paths encircling a moiré cell[27], comprising an *AB* and *BA* triangle. The discrepancy between the experiment and theory is attributed to the ambiguity in the determination of the twist angle because the lack of spatial resolution in transport techniques[27]. The high spatial resolution of the STM can naturally overcome this limitation and accurately measure the area encircled by the paths. In this work, not only the magnetic AB oscillations, but also the electrostatic AB effect in a moiré-scale triangular quantum path in the minimally TBG are directly measured by using the STM.

In our experiment, the TBG were obtained by transferring layer by layer of the large-area aligned monolayer graphene onto a Nb-doped $SrTiO_3$ (0.7%) substrate with controlled twist angles, as reported in ref.[16] (see Supplementary Section 1 for details). Different TBG with various twist angles were obtained and in this work we focused on the minimally TBG with $\theta < \theta_c$, *i.e.*, the TBG with a triangular network of the DWs[14,16] (see Supplementary Section 2 for details). Figures 2a-2c show representative STM images of three TBG with different twist angles, $\theta = 0.18°$, $\theta = 0.61°$ and $\theta = 0.88°$. Twist-induced varied stacking regions of the TBG are also characterized at atomic level further. For example, the well-defined transition from triangular contrast lattice to hexangular-like lattice and then to triangular contrast lattice is observed across the DWs in atomic-resolved STM images (see Supplementary Sections 3 for details). For the TBG with $\theta < \theta_c$, the competition between the interlayer van der Waals coupling and the intralayer elastic deformation leads to structural reconstruction. Consequently, the *AB* and *BA* stacking regions are enlarged and the *AA* stacking regions are reduced to minimize the total energy of the system[14]. A clear signature of the lattice reconstruction in the TBG can be clearly identified by the triangular network of DWs connecting the *AA* stacking regions, as shown in the STM images of all the three TBG (Figs. 2a-2c). To further characterize the TBG, we also carried out scanning tunneling spectroscopy (STS) measurements in different stacking regions (see Supplementary Sections 3-5 for details). In the *AA* stacking regions, a pronounced DOS peak, which arises from the low-energy flat bands in the minimally TBG[28,29], is obtained. In the *AB* and *BA* stacking regions, a finite gap, which is generated by the inequivalent charge densities of the two topmost adjacent layers[22-24,30-33], is observed in the tunneling spectra. In the meanwhile, the symmetry-protected gapless mode, as guaranteed by opposite valley Chern number of the gapped *AB* and *BA* stacking regions, emerges in the DWs. The triangular network of the 1D conducting channels along the DWs are directly imaged in all the three TBG by operating energy-fixed STS mapping at energies within the gap of the Bernal stacking regions (see Supplementary Sections 3-5 for details). Below, we will show that electronmagnetic potentials can modulate the electron interference in paths along the triangular 1D conducting channels and, consequently, result in oscillations of the local DOS in the DWs, which can be interpreted as due to the AB effect.

In the STS measurements, the value of the spectra at a fixed bias is proportional to the local DOS at a selected energy and the spectra directly reflect the local DOS beneath the STM tip. Therefore, we can obtain the local DOS modulated by the magnetic fields by measuring the tunneling spectra (see Supplementary Section 4 for STS measurements under magnetic fields). Figures 2d-2f show representative values of the spectra at a fixed bias as a function of magnetic fields recorded at the DWs of the three TBG. There is no observable DOS oscillation for the 0.88° TBG and only very weak DOS oscillation for the 0.61° TBG (see Supplementary Section 5 for details of the STS spectra). However, pronounced DOS oscillations are clearly observed for the 0.18° TBG and are attributed to the AB effect for electrons propagating along the triangular network of the 1D conducting channels hosted by the DWs. Such a result is quite reasonable since that the size of each Bernal stacking domain in the 0.88° and 0.61° TBG is relative small, which makes that the Bernal stacking domain is not well-defined insulating. Therefore, the electron interference modulated by magnetic flux in paths encircling a Bernal stacking domain is strongly suppressed. Additionally, the main periodicities $\Delta B$ for the 0.88° TBG and 0.61° TBG are about 37 T and 18 T respectively, which are much larger than the maximum field applied in our experiment (the main periodicities are calculated with assuming that the AB oscillations arise from the interference in paths encircling a *AB* or *BA* domain). This may also weak the DOS oscillations induced by the AB effect in the 0.88° TBG and the 0.61° TBG.

There are other possible origins, such as Shubnikov-de Haas (SdH) oscillations[34,35], that can lead to oscillations of the DOS modulated by magnetic fields. Usually, the period of these oscillations is sensitive to the measured energy, which is quite different from the AB effect with the period that is independent of the energy. To further explore the origin of the observed DOS oscillations in the 0.18° TBG, we measured the DOS oscillations at different energies, as shown in Fig. 3a. Obviously, the periodicities of the DOS oscillations are almost independent of the energy, further confirming that the oscillations arise from the AB effect. In the 0.18° TBG, there is a large heterostrain that strongly modifies the local period of the morié superlattice[17,18,36,37], as schematically shown in Fig. 3b, which strongly affects the observed AB oscillations. With assuming that the oscillations arise from the AB effect, the expected periods of magnetic field $\Delta B$ can be roughly estimated as 2.58 T, 1.97 T, 1.18 T, and 0.66 T for electrons propagating along four different paths around the measured position (Fig. 3b). It is interesting to note that the calculated values of the $\Delta B$ are consistent with the observed AB resonances in our experiment (Fig. 3a), further confirming that the DOS oscillations arise from the magnetic AB effect. The advantage of our experiment is the ability to obtain the area encircled by the quantum paths and the periods of magnetic field simultaneously. According to our experiment, the main period of the oscillations should be attributed to interference in paths encircling a *AB* or *BA* domain, rather than a morié pattern.

The high-spatial resolution of the STM enables us to measure the AB effect at the nanoscale. Figure 4 shows representative DOS oscillations measured at different positions along a DW in the 0.18° TBG. Surprisingly, the maxima of the oscillations shift on different measured positions, which is unexpected when only the magnetic AB effect is taking into account. According to the result in Fig. 4, a maximum DOS (constructive interference) is turned into a minimum DOS (destructive interference) either by changing the magnetic fields or by changing the measured positions. This symmetry reminds us that the shift of the maxima on different positions arises from the electrostatic potential, which also can change the phase of electron wave function[1,5,6].

The different stacking regions and/or the substrate can generate a finite in-plane electrostatic potential along the paths of electrons in the minimally TBG. An electrostatic potential difference of about 20 mV can lead to the observed electrostatic AB effect, as shown in Fig. 4 (see Supplementary Section 6 for details). Such a value of the potential difference is roughly consistent with the measured variation of charge neutrality points in the TBG according to the STS spectra.

In conclusion, we demonstrated for the first time that STM can be used to study the AB effect at the nanoscale. Thanks to the high spatial resolution of the STM, both the magnetic and electrostatic AB effects are measured in a moiré-scale triangular quantum path in the minimally TBG.

**Acknowledgements**

This work was supported by the National Natural Science Foundation of China (Grant Nos. 11974050, 11674029). L.H. also acknowledges support from the National Program for Support of Top-notch Young Professionals, support from "the Fundamental Research Funds for the Central Universities", and support from "Chang Jiang Scholars Program".

# Figures

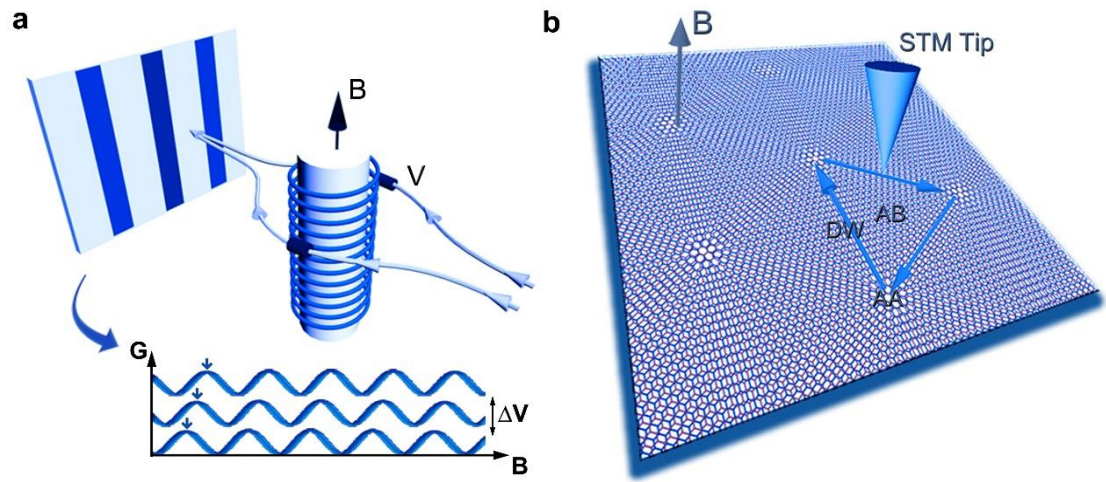

**Figure 1** | Schematic of Aharonov-Bohm effects realized in experiments. **a,** There is a phase shift for the coherent electron beams along two trajectories after applying external electromagnetic fields (top panel). The periodic conductance fluctuations as a function of electromagnetic fields can be detected in transport experiments and the peaks of oscillations are marked by arrows (bottom panel). **b,** Schematic of the STM for measuring the AB oscillations in a TBG. AA, AB, DW label different stacking regions in the TBG. The DWs can form a closed loop for electron interference (bule arrows). The STM tip acts as a detector to measure the electron-wave interference in a morié-scale path.

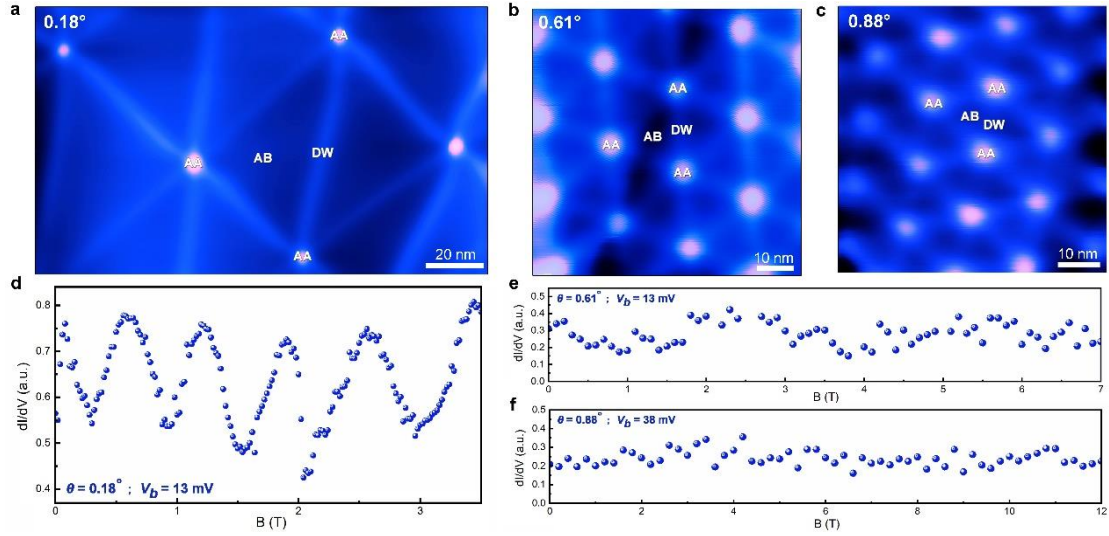

**Figure 2** | Local DOS oscillations modulated by magnetic fields in the TBG with different twist angles. **a-c,** STM topographic images of the morié superlattices with twist angle $\theta \sim 0.18°$ ($V_b$ = 700 mV, $I$ = 100 pA), $\theta \sim 0.61°$ ($V_b$ = 300 mV, $I$ = 400 pA), and $\theta \sim 0.88°$ ($V_b$ = 300 mV, $I$ = 400 pA). The AA, AB, and DW stacking orders in the TBG are marked on the corresponding regions. **d-f**, *dI/dV* signal at a fixed tunneling bias as a function of the magnetic fields in the TBG with different twist angles. The oscillations become obvious in the $\theta = 0.18°$ TBG.

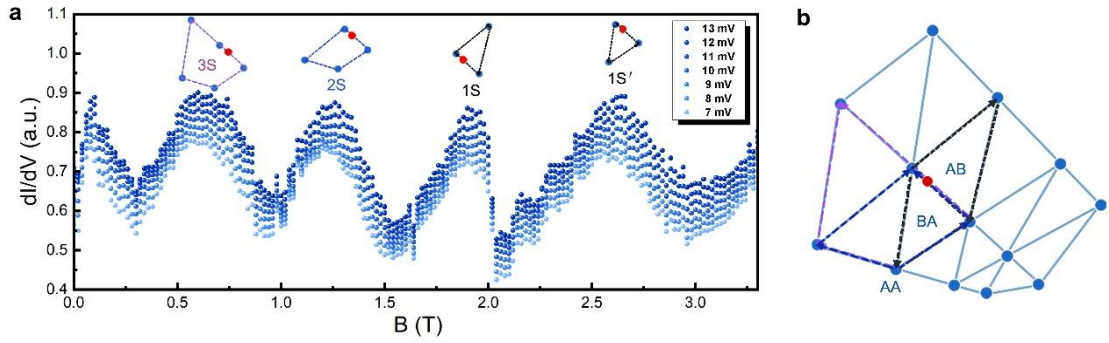

**Figure 3** | Magnetic Aharonov-Bohm oscillations in the 0.18° TBG at different energies. **a,** *dI/dV* signal at various tunneling bias as a function of the magnetic fields recorded at a fixed position of a DW in the 0.18° TBG. The maxima of the *dI/dV* oscillations are at about 0.6 T, 1.2 T, 1.9 T and 2.6 T. **b,** Schematic of the network of the DWs in experiment. The STS spectra are recorded at the position marked by the red dot. Colored dashed lines with different colors label four possible paths for electron interferences due to the AB effect. The maxima of the oscillations are attributed to interference in the four paths due to the magnetic AB effect, as schematically shown in panel a.

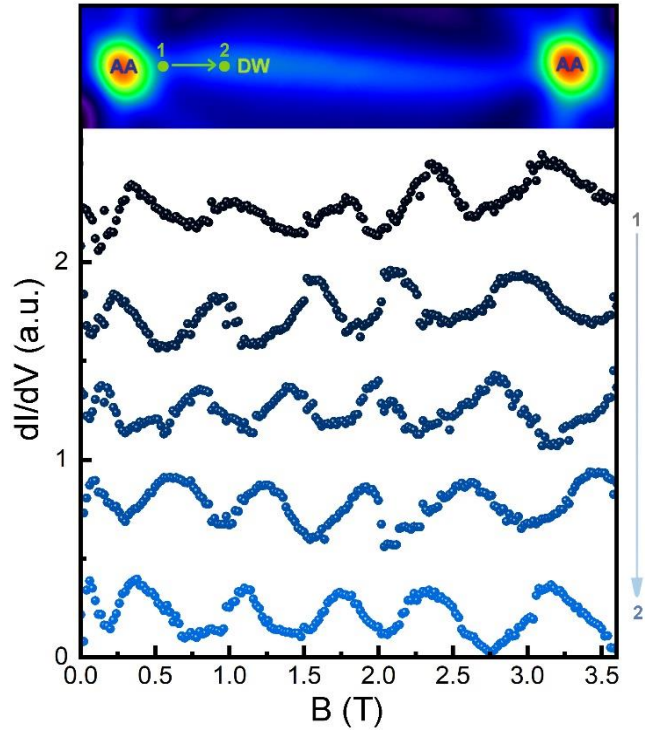

**Figure 4 | Magnetic and Electrostatic Aharonov-Bohm effect in the 0.18° TBG.** *dI/dV* signal at a fixed tunneling bias as a function of the magnetic fields recorded at different positions, from position 1 to position 2 in the upper panel, of a DW in the 0.18° TBG. At a selected position, the *dI/dV* signal varies and exhibits maxima as a function of magnetic fields. At a fixed magnetic field, the maximum of the *dI/dV* signal is turned into a minimum by changing the measured positions, attributing to the electrostatic AB effect.